\documentclass[11pt]{article}
\pdfoutput=1
\usepackage[margin=1.0in]{geometry}
\usepackage{amsmath,amssymb,graphicx}
\usepackage{hyperref}
\usepackage{slashed}

\usepackage[T1]{fontenc}
\usepackage[p]{cochineal}
\usepackage[varqu,varl,var0]{inconsolata}
\usepackage[scale=.95,type1]{cabin}
\usepackage[cochineal, vvarbb]{newtxmath}
\usepackage[cal=boondoxo]{mathalfa}

\usepackage[utf8]{inputenc}

\usepackage{upgreek}

\usepackage[font=small,labelfont=bf]{caption}
\usepackage{cite}

\newcommand{\iu}{{\mathrm i}}
\newcommand{\E}{{\mathrm e}}

\newcommand{\dif}[1]{\ensuremath{\operatorname{d}\!{#1}}}

\newcommand{\be}{\begin{equation}}
\newcommand{\ee}{\end{equation}}

\usepackage{xcolor}
\usepackage{bm}

\title{\bf  Photon Masses in the Landscape and the Swampland}
\author{Matthew Reece\\
{\small \color{gray} \texttt{mreece~(@physics.harvard.edu)}}\\
{\small Department of Physics, Harvard University, Cambridge, MA, 02138}}

\begin{document}
\maketitle

\begin{abstract}
In effective quantum field theory, a spin-1 vector boson can have a technically natural small mass that does not originate from the Higgs mechanism. For such theories, which may be written in St\"uckelberg form, there is no point in field space at which the mass is exactly zero. I argue that quantum gravity differs from, and constrains, effective field theory: arbitrarily small St\"uckelberg masses are forbidden. In particular, the limit in which the mass goes to zero lies at infinite distance in field space, and this distance is correlated with a tower of modes becoming light according to the Swampland Distance Conjecture. Application of Tower or Sublattice variants of the Weak Gravity Conjecture makes this statement more precise: for a spin-1 vector boson with coupling constant $e$ and St\"uckelberg mass $m$, local quantum field theory breaks down at energies at or below $\Lambda_{\rm UV} = \min((m M_{\rm Pl}/e)^{1/2}, e^{1/3} M_{\rm Pl})$. Combined with phenomenological constraints, this argument implies that the Standard Model photon must be exactly massless. It also implies that much of the parameter space for light dark photons, which are the target of many experimental searches, is compatible only with Higgs and not St\"uckelberg mass terms. This significantly affects the experimental limits and cosmological histories of such theories. I explain various caveats and weak points of the arguments, including loopholes that could be targets for model-building.
\end{abstract}

\section{Introduction}

\subsection{The case for a photon mass in effective field theory}

Many of the biggest mysteries of the Standard Model are associated with small numbers. Some are measured to be nonzero, but their size is difficult to explain in terms of fundamental scales (e.g.~the cosmological constant, the Higgs boson mass). Some are so far measured to be consistent with zero, despite every expectation that they are actually nonzero (e.g.~the strong CP phase $\overline \theta$). One small parameter that we usually don't discuss in the latter category is the mass of the photon. Why do we not view it as a mystery? Like the cosmological constant, it might be puzzlingly small, but nonzero. An unsatisfactory answer, which puts the cart before the horse, is that the photon must be massless because of gauge invariance. A theory of a massless spin-1 particle necessarily has a gauge redundancy when written in terms of local quantum fields. But the theory of a photon with a very small mass is perfectly consistent, with no gauge redundancy \cite{Proca:1900nv}. To declare that gauge invariance is fundamental is simply to assert that the photon should be massless. This is not a very satisfying answer. 

A massless photon has two independent physical degrees of freedom (helicity states) whereas a massive one has three; in this sense, the $m_\gamma \to 0$ limit is discontinuous. However, both cases are consistent with data, because the longitudinal mode of the photon couples very weakly to matter when the photon mass is small \cite{bass1955must}. As Sidney Coleman explained, the correct response to the notion that we can determine whether the photon is massive by counting degrees of freedom using the Planck law for a hot oven is ``This is {\em garbage}.... it will require twenty trillion years for that oven to reach thermal equilibrium!''~\cite[\S26.4]{Chen:2018cts}. Data can at best set an upper bound on the photon mass, so the cases with $m_\gamma = 0$ (with two degrees of freedom) and $m_\gamma \neq 0$ (with three degrees of freedom) are equally viable in an effective quantum field theory description of our universe. The situation is quite closely analogous to that of neutrino masses. For many years it was possible to believe that neutrinos were exactly massless Weyl fermions. In light of data, we now know that neutrino masses are nonzero. However, two possibilities are equally consistent with all current data: neutrinos have Majorana masses, represented as the higher dimension operator $(h \cdot \ell)^2$ in the Standard Model; or they have Dirac masses, $(h \cdot \ell) N$, where new fermions $N$ that are Standard Model singlets must be added to the theory. We do not view the former theory as a more likely description of our universe simply because it incorporates fewer degrees of freedom. Rather, until we acquire better data on lepton-number violation we must consider these as two, equally valid, possible Standard Models. By analogy, we could consider two different variations of the Standard Model, one with a massive photon (including an additional degree of freedom) and the additional terms that are allowed once we give up gauge invariance, such as $-\lambda_\gamma (A_\mu A^\mu)^2$. Only empirical evidence can settle the question.

In light of the analogous cases of the cosmological constant and neutrino masses, which were believed to be exactly zero by some physicists until they were empirically demonstrated to be nonzero, it is tempting to suggest that we should believe that the photon is {\em likely} to have a tiny but nonzero mass, there being no field-theoretic prejudice against it. This point of view has been expressed before, for instance in \cite{Heeck:2013cfa}, which argues that there is no theoretical prejudice in favor of $m = 0$, sets a bound on the photon lifetime, and explicitly argues that the smallness of the photon mass is similar to the strong CP problem. 

The simplest experimental bounds on the photon mass are purely kinematic. They measure the dispersion in arrival times of photons of different frequencies emitted by Fast Radio Bursts (FRBs) to set stringent constraints: $m_\gamma \lesssim 2 \times 10^{-14}~{\rm eV}$~\cite{Wu:2016brq, Bonetti:2016cpo, Bonetti:2017pym}. Because radio photons have long wavelengths, such constraints have a significant kinematic advantage over those from other astrophysical processes like Gamma Ray Bursts. Aside from these straightforward kinematic tests, there is a huge range of additional experimental results, reviewed in \cite{Goldhaber:2008xy}. A bound $m_\gamma \lesssim 6 \times 10^{-16}~{\rm eV}$ arises from measurements of Jupiter's magnetic field \cite{Davis:1975mn}. Even stronger bounds have been claimed from solar system magnetic fields \cite{Ryutov:2007zz, Ryutov:2009zz} and from galactic magnetic fields \cite{Chibisov:1976mm}. However, these bounds require more modeling assumptions, both in terms of the structure of magnetic fields \cite{Retino:2013gga} and in whether the underlying theory of photon mass permits the existence of vortices \cite{Adelberger:2003qx}. Below I will mostly take the FRB constraint as a benchmark, though in some cases my conclusions could be strengthened by considering other bounds.

Theories of massive photons admit additional new interactions, absent in the usual Standard Model. For instance, we could consider self-interactions, such as $-\lambda_\gamma (A_\mu A^\mu)^2$. To the best of my knowledge, despite the huge literature constraining photon masses, there has been little work on constraining such non-gauge-invariant photon interactions. To set a crude constraint on such an operator, we can adapt an argument of \cite{Liang:2011sj}: the existence of visible light reaching us from distant stars, despite the presence of numerous CMB photons from which photons could scatter, tells us that the mean free path $\ell$ of a photon in our universe is very long. If we estimate that $\ell \sim E_\gamma/(T_{\rm CMB}^2 \lambda_\gamma^2) \gtrsim H^{-1}$, we infer $\lambda_\gamma \lesssim 10^{-13}$. Considering radio photons can improve this by a few orders of magnitude. This appears to be a strong constraint. However, one should be careful in making such arguments: because the longitudinal polarization vector appears with a factor of $E/m_\gamma$, scattering through such self-interactions will violate perturbative unitarity above some low energy, and a UV completion is needed to make sense of what it means to set a bound on this theory. Some discussions of effective field theories with massive photons with nonminimal self-couplings or gravitational couplings appear in \cite{Heisenberg:2014rta, Jimenez:2016isa, Chowdhury:2018nfv}.

%This is a strong constraint; whether it is more or less constraining than the bound on $m_\gamma$ requires a more complete theory of the origin of the photon mass.

In summary, phenomenological constraints imply that the Standard Model photon mass and any associated interactions that violate gauge invariance must be very small. However, within the context of effective quantum field theory, there is no strong argument that they should be exactly zero. Indeed, a small photon mass is {\em technically natural} in the sense that radiative corrections do not generate a much larger mass. The particle theory community has recently been very open-minded about considering theories with extremely tiny unexplained parameters, provided they are technically natural. A careful effective field theorist, then, should augment the usual Standard Model with terms like $\frac{1}{2} m_\gamma^2 A_\mu A^\mu - \lambda_\gamma (A_\mu A^\mu)^4 + \cdots$ and constrain these terms experimentally.

\subsection{The case against a photon mass in quantum gravity}

Everything I have just told you is ridiculous. You don't believe that the Standard Model photon mass is nonzero. I certainly don't.

The challenge is to put our instinct that a nonzero photon mass is ridiculous on a firmer footing. Our arguments have been correct as far as they go; however, they are limited by being based solely on effective quantum field theory. Our universe has a very important property that motivates going beyond effective quantum field theory, namely, gravity. It is well known that many phenomena that appear innocuous in effective field theory are forbidden in theories of quantum gravity: examples include continuous global symmetries \cite{Banks:1988yz,Kamionkowski:1992mf,Holman:1992us,Kallosh:1995hi,Banks:2010zn} or abelian gauge groups without quantized charges \cite{Banks:2010zn}. These examples motivate us to frame the question of whether a very light St\"uckelberg vector boson mass is allowed as a question about the boundary between the landscape of consistent theories of quantum gravity and the so-called ``Swampland'' of effective field theories without UV completions \cite{Vafa:2005ui,Arkanihamed:2006dz,Ooguri:2006in,Brennan:2017rbf}.

Our arguments apply to St\"uckelberg masses but not to the Higgs mechanism. It is important, then, that we give a clear explanation of the relevant distinction between the two mechanisms. In the Higgs mechanism, we have a scalar field $\phi$ of charge $q$ which obtains a vacuum expectation value $\langle \phi \rangle = \frac{1}{\sqrt{2}} v$, so that the kinetic term $|D_\mu \phi|^2 = |\partial_\mu \phi - \iu e q {\hat A}_\mu \phi|^2$ (written in terms of the canonically normalized gauge field ${\hat A}$) yields an effective mass $m_A = e q v$. The field $\phi$ decomposes into a radial excitation $\sigma$, which in general is massive, and an azimuthal Goldstone mode $\theta$ which is eaten to provide the longitudinal mode of the gauge field. The charge $q$ is quantized, because the gauge group in a theory coupled to gravity must be U(1) and not $\mathbb{R}$. Hence the mass $m_A$ approaches zero in two limits: either $e \to 0$ or $v \to 0$. The limit $e \to 0$ leads to a continuous global symmetry, which is forbidden in quantum gravity. A sharper version of this statement is given by the Weak Gravity Conjecture (WGC) and its variants, which imply two sorts of UV cutoffs as $e \to 0$: a ``Magnetic WGC cutoff'' at or below $e M_{\rm Pl}$ \cite{Arkanihamed:2006dz}, which in known examples is associated with the appearance of a tower of charged states that must be included in the theory \cite{Heidenreich:2015nta,Heidenreich:2016aqi,Montero:2016tif,Andriolo:2018lvp}, and a more severe quantum gravity cutoff at or below $e^{1/3} M_{\rm Pl}$ at which local field theory breaks down entirely due to the net quantum effects of particles in the tower \cite{Heidenreich:2016aqi,Heidenreich:2017sim}. However, taking $m_A$ small using the $v \to 0$ limit is unproblematic. It may be fine-tuned, in a theory of an elementary Higgs boson, but more generally if $\phi$ is a composite operator (as in dynamical symmetry breaking by QCD-like dynamics) we can naturally obtain a small mass $m_A$. For our present purposes, such dynamical symmetry breaking is considered a special case of the Higgs mechanism.\footnote{This is a mild over-simplification; after reading the argument below the interested reader can consult Appendix~\ref{app:DSB} for a more careful statement about the case of dynamical symmetry breaking.}

St\"uckelberg theories can be obtained from the Proca Lagrangian by introducing a fictitious gauge symmetry~\cite{Stueckelberg:1900zz}. We have a scalar field $\theta$ with the gauge transformation properties ${\hat A}_\mu \mapsto {\hat A}_\mu + \frac{1}{e} \partial_\mu \alpha$, $\theta \mapsto \theta + \alpha$, and the photon mass term is $\frac{1}{2} f_\theta^2 (\partial_\mu \theta - e {\hat A}_\mu)^2$. Then the photon mass is $m_A = e f_\theta$, and it may be made small in two different limits: $e \to 0$ or $f_\theta \to 0$. Superficially, this is not so different from the case discussed above: the Higgs field VEV $v$ has been replaced with the parameter $f_\theta$, and the radial excitation $\sigma$ is absent from the theory. At first glance one might take the absence of a radial excitation to be the defining feature of St\"uckelberg, as opposed to Higgsed, massive vector bosons. However, this is {\em not} the correct definition for our purposes. In particular, in any theory with supersymmetry (broken or not), a radial excitation always appears in the theory that is loosely analogous to the Higgs boson $\sigma$. This is because supersymmetry packages the field $\theta$ into a complex scalar multiplet containing two real fields. Specifically, we have a vector superfield $A$ and a chiral superfield $\Phi$ for which a U(1) gauge transformation sends $A \mapsto A + \iu ( \alpha - \alpha^\dagger )$ and $\Phi \mapsto \Phi - \iu c \alpha$, so that the K\"ahler potential for $\Phi$ must have the form $K(\Phi + \Phi^\dagger + c A)$. The imaginary part of $\Phi$ plays the role of the field $\theta$, and its real part plays the role of the radial excitation $\sigma$. If the salient difference between St\"uckelberg and Higgsing were only the radial mode, one might initially suspect that there is no difference at all between the two mechanisms in the supersymmetric context.

Nonetheless, there is a well-defined physical distinction between the Higgs mechanism and the St\"uckelberg mechanism. The key difference is that in the Higgs case the value $\phi = 0$ at which $m_A = 0$ is a well-behaved point in field space, whereas in the St\"uckelberg case it is not. In the minimal non-SUSY St\"uckelberg mechanism, there is no radial mode and the mass is a fixed quantity. In the SUSY context, or more generally, the radial mode may exist, but it never attains a value corresponding to $m_A = 0$. The massless limit lies at infinite distance in field space. In other words, the kinetic term for the radial mode is singular at the point where the vector boson mass is zero. This means that we can only distinguish between St\"uckelberg and Higgs theories once we know the form of the kinetic terms, i.e.~of $K(\Phi + \Phi^\dagger + c A)$ in the SUSY context.

Indeed, St\"uckelberg masses in this sense are ubiquitous in string theory \cite{Ibanez:2012zz, Quevedo:2016tbh}, and their common feature is not the absence of a radial mode but rather that the vector boson mass is nonzero throughout the entire field space. The classic example has a K\"ahler potential of the form \cite{Dine:1987xk}
\be
K(\Phi, \Phi^\dagger, A) = -M^2 \log(\Phi + \Phi^\dagger + c A),
\ee
so that the $m_A \to 0$ limit is $\Phi \to \infty$. The field-space distance to a point of small mass behaves as $|\log(m_A^2)|$. That this diverges as $m_A \to 0$ is the most important aspect of the theory for our argument.\footnote{St\"uckelberg gauge field masses in string theory are often, though not always, associated with 4d Green-Schwarz anomaly cancelation \cite{Dine:1987xk,Aldazabal:1998mr, Ibanez:1998qp, Ibanez:1999pw, Ibanez:1999it, Antoniadis:2002cs}. In some cases they are linked to 6d anomalies \cite{Antoniadis:2002cs,Anastasopoulos:2003aj,Anastasopoulos:2006cz}. The association of St\"uckelberg masses with anomalies in quantum gravity theories will play no role in the arguments of this paper, but would be interesting to explore further. It may provide refinements of, or additional arguments for, the conjectures discussed in this paper.}

At this stage, the main argument of this paper will be apparent to readers familiar with the Swampland literature. Known quantum gravity theories have a common feature which motivates the Swampland Distance Conjecture: when venturing over large distances in field space, a tower of states will become light, eventually invalidating the low-energy effective quantum field theory description \cite{Ooguri:2006in, Baume:2016psm, Klaewer:2016kiy, Blumenhagen:2017cxt, Cicoli:2018tcq, Heidenreich:2018kpg, Grimm:2018ohb}. This appears to be related to a ``trans-Planckian censorship'' phenomenon in general relativity, which makes it impossible for observers to probe very large scalar field ranges \cite{Nicolis:2008wh, Dolan:2017vmn}. This suggests that in the St\"uckelberg context, unlike the Higgs context, the limit $m_A \to 0$ will be associated with a breakdown of effective quantum field theory. The reflection of this fact at finite but small $m_A$ is the existence of an ultraviolet cutoff which tends to zero as $m_A \to 0$ (but which will disappear as $M_{\rm Pl} \to \infty$). These arguments suggest that if we start with a St\"uckelberg theory for which $m_A = e f_\theta$, any attempt to take small $m_A$ through the route of small $f_\theta$ will lead to a breakdown of effective quantum field theory at a UV cutoff energy
\be
\Lambda_{\rm UV} \lesssim f_\theta^\alpha M_{\rm Pl}^{1-\alpha},   \label{eq:generalalpha}
\ee
for some constant $0 < \alpha < 1$. Below, we will motivate $\alpha = 1/2$ as the correct version of this conjecture.

\subsection{Outline}

\begin{figure}[tp]
\centering
\includegraphics [width = 0.5\textwidth]{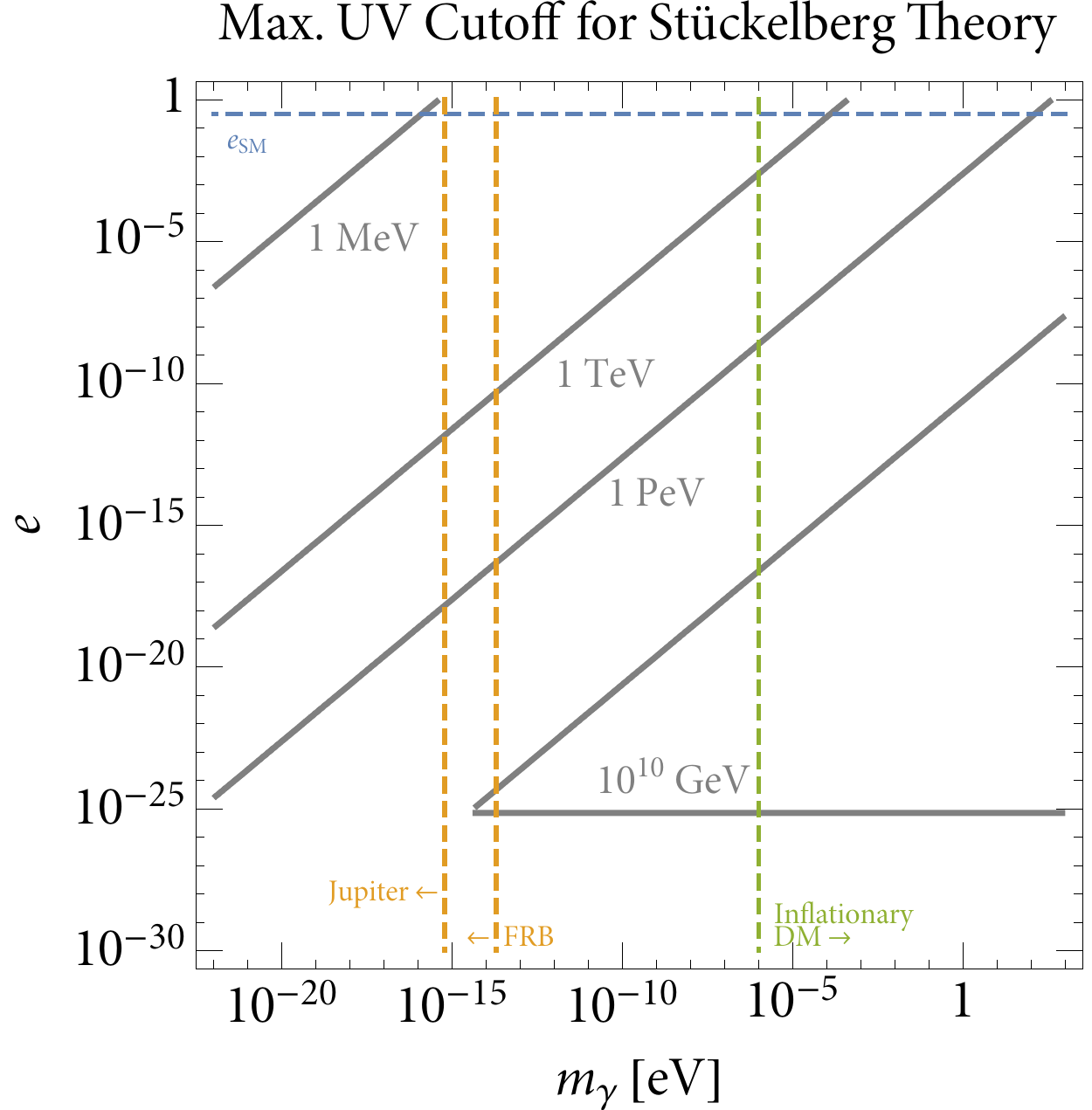}
\caption{
Contours of the maximum possible UV cutoff on a theory of a St\"uckelberg photon, as a function of the photon mass $m_\gamma$ and the gauge coupling $e$. Unless $e$ is exceptionally tiny, the bound is set by $\sqrt{m_\gamma M_{\rm Pl}/e}$. Vertical orange dashed lines depict upper bounds on the Standard Model photon mass: $6 \times 10^{-16}~{\rm eV}$ from Jupiter's magnetic field \cite{Davis:1975mn}; $2 \times 10^{-14}~{\rm eV}$ from FRBs \cite{Wu:2016brq, Bonetti:2016cpo, Bonetti:2017pym}. The vertical green dashed line marks the approximate smallest mass for which generation of dark photon dark matter from inflationary fluctuations is efficient \cite{Graham:2015rva}. 
}
\label{fig:UVcutoff}
\end{figure}

We have already sketched the central argument of this paper. The defining distinction between vector boson masses of St\"uckelberg and Higgs type is whether the massless point in field space lies at infinite distance or finite distance. Applying existing Swampland conjectures related to infinite distances in field space then immediately implies that very small St\"uckelberg masses are sick. In the remainder of the paper, we will formulate a more precise version of this conjecture, which leads to the general bound
\be
\Lambda_{\rm UV} \lesssim \min((m M_{\rm Pl}/e)^{1/2}, e^{1/3} M_{\rm Pl}).
\ee
This bound is illustrated in Fig.~\ref{fig:UVcutoff}. 

We will begin in \S\ref{sec:axion} by discussing a closely related bound on theories of pure axions, then argue in \S\ref{sec:stuckelberg} that the bound generalizes to St\"uckelberg vector bosons. We will show that our proposed bounds are satisfied by familiar examples of axions and St\"uckelberg vector bosons in string theory. Next, we will discuss phenomenological applications of this bound in \S\ref{sec:pheno}. The first is to the question of whether the Standard Model photon has a nonzero mass; our arguments strongly suggest that it does not. The second is to possible massive ``dark photons,'' as-yet-undiscovered light vector bosons, in our universe. We will see that much of the parameter space probed by current or proposed future experiments searching for dark photon dark matter is incompatible with St\"uckelberg masses, in light of our arguments. Our arguments do not apply to the Higgs mechanism, which can also provide a mass to the dark photon. However, the cosmology and phenomenology of a dark photon obtaining a mass from the Higgs mechanism can differ from the pure St\"uckelberg case, so this conclusion may have important consequences for dark photon modeling. Some further comments and clarifications on the arguments of the paper are given in \S\ref{sec:comments}. We conclude in \S\ref{sec:conclusions} with some general remarks on prospects for putting the conjectures on a more solid footing and broader implications for the notion of naturalness.

I should emphasize that the arguments in this paper are not rigorous, even by the relaxed standards of a ``physics proof.'' They rely on various existing conjectures, and in some cases require extensions of these conjectures to be true. It is worth briefly summarizing various caveats up front. First, as should be clear from the above discussion, our arguments do not apply to photon masses arising from the Higgs mechanism. If the quantization of electric charge allows tiny fractional charges, then a Higgs with such a charge could produce a consistent photon mass, providing one loophole to our arguments. In the St\"uckelberg case (defined as above), we will mostly assume that the structure of quantum gravity still requires underlying compact gauge groups; some of our arguments assume that the Weak Gravity Conjecture generalizes to this setting (despite the presence of masses). We will sketch some arguments for why these assumptions are plausibly true, and how a version of the argument might survive when relaxing the compactness assumption. Because our argument builds on the Swampland Distance Conjecture, it is worth noting that this conjecture is known to be true in many string theory settings. Our argument applies directly to those theories, where one might otherwise have carried out a case-by-case analysis. However, the Swampland Distance Conjecture has not been proven from general principles. The most quantitative version of our bound relies on identifying the energy scale of a string tension with a UV cutoff. Making this argument completely precise is somewhat tricky. Furthermore, even if the quantitative version of our bound as stated can be proven for a single massive photon, we have not studied how it generalizes to a theory with multiple massive photons. It is possible that a mass matrix whose entries are bounded as we discuss can nonetheless have a much smaller eigenvalue. Each of these caveats (discussed at greater length below) could serve as a starting point for further work, either as a direction in which to seek counterexamples or as a place to shore up the foundations and strengthen the argument.

\section{Fundamental axions and UV cutoffs}
\label{sec:axion}

We can think of a vector boson with a St\"uckelberg mass as a gauge field that has eaten a compact (periodic) scalar axion field. (We will comment below on the viability of the alternative case with a noncompact scalar field.) Before discussing the St\"uckelberg case, then, we should first discuss a simpler question: can compact axion fields in quantum gravity have arbitrarily small period? I will argue that in the case of a {\em fundamental} axion field, we expect a small period (i.e., decay constant) to be associated with a low ultraviolet cutoff on the theory.

Here I draw a distinction between a {\em fundamental} axion, of which closed string axions \cite{Witten:1984dg,Choi:1985je,Barr:1985hk,Conlon:2006tq,Svrcek:2006yi} or more generally Wilson lines of gauge fields around extra-dimensional cycles \cite{ArkaniHamed:2003wu} are examples, and an ordinary field-theoretic (pseudo-)Nambu-Goldstone boson, of which open string axions \cite{Berenstein:2012eg,Honecker:2013mya} are examples. In both cases, in the quantum gravity context we expect the axion to be accompanied by a radial mode (see the discussion of Conjecture 4 in \cite{Ooguri:2006in}), which in the supersymmetric context is a ``saxion.'' The distinction between the two varieties of axion is precisely analogous to the one we have discussed earlier for the St\"uckelberg and Higgs mechanisms: for an ordinary Nambu-Goldstone boson there is a point in field space at which the decay constant vanishes, whereas for a fundamental axion such a point lies at infinite distance. An open string axion is simply the phase of some weakly-coupled complex scalar field, with a well-behaved kinetic term at the origin. Contrast this with the case of a closed string axion arising from a $p$-form gauge field in extra dimensions integrated over a $p$-cycle:
\be
\theta = \int_{\Sigma_p} C_p.
\ee
The radial partner $\tau$ is generally related to the volume of $\Sigma$; in familiar examples arising in string or Kaluza-Klein theory, we find that the kinetic terms behave as
\be
{\cal L}_{\rm kin} \sim \frac{M_{\rm Pl}^2}{\tau^2} \left(\partial_\mu \theta \partial^\mu \theta + \partial_\mu \tau \partial^\mu \tau\right).
\ee 
The important distinction from a standard Goldstone boson picture is that {\em no point in field space exists where the period of the axion vanishes}, because this corresponds to an infinite volume limit $\tau \to \infty$. It lies a logarithmically infinite distance away in the field space metric. Hence, Wilson loop (or surface) axion fields are examples of what I refer to as fundamental axions.

\subsection{Conjectures for axions}

Now that we have introduced our basic concepts and terminology, we will formulate two conjectures for axions in theories of quantum gravity.

\bigskip
\noindent
{\em Radial Mode Mass Bound}\footnote{My remarks about this bound are heavily indebted to discussions with Thomas Dumitrescu, though any shortcomings in the version presented in this paper are my own.}

Ooguri and Vafa claimed that any periodic axion field $\theta(x)$ in a theory of quantum gravity will be accompanied by a radial mode $\sigma(x)$ (see Conjecture 4 of \cite{Ooguri:2006in} and its accompanying discussion). By ``radial mode'', we mean that a coupling of the form $\sigma (\partial \theta)^2$ exists in the theory. In a supersymmetric theory, this mode is the ``saxion,'' which in the limit of unbroken SUSY must be massless whenever the axion is massless.

In the context of an ordinary, weakly-coupled Goldstone boson, it is difficult to arbitrarily decouple the radial mode. For example, if we have a theory with a symmetry breaking potential $\lambda(|\phi|^2 - \frac{1}{2} f_\theta^2)^2$, then we cannot make the radial mode of $\phi$ heavier than about $4\pi f_\theta$ without violating the perturbative unitarity bound on the coupling $\lambda$. In the case of a fundamental axion, such as a closed string axion, the situation is somewhat less clear, since the effective theory is more complicated (the kinetic term for $\phi$ is not well-behaved at $\phi \to 0$). However, it still seems likely that a bound exists and that the radial mode cannot be arbitrarily decoupled.

A precise conjecture is that an axion of period $2 \pi f_\theta$ in a theory of quantum gravity must be accompanied by a ``saxion'' or radial partner obeying a bound
\be
m_{\sigma} \lesssim 4 \pi f_\theta.
\ee
We can argue for this bound using instanton effects. We expect that any axion in a quantum gravity theory couples to some instantons that explicitly break its continuous shift symmetry to a discrete one; for example, this follows from a variant of the WGC \cite{Arkanihamed:2006dz}. (More generally, we expect similar effects to exist in most interacting QFTs, not just theories of quantum gravity.) These instantons do not necessarily generate an axion potential, as they may be accompanied by other factors (e.g.~fermion zero modes). Nonetheless, we expect that the effective action contains terms like ${\cal O} \E^{\iu a / f_\theta} + {\rm h.c.}$ where $a = f_\theta \theta$ is the canonically normalized axion, for some (possibly complicated) operator $\cal O$. We can then draw diagrams where we close up axion lines emerging from this operator into quadratically divergent loops, producing correction terms multiplied by $\Lambda_{\rm UV}^2 / (4\pi f_\theta)^2$ factors. In order for loops not to dominate over the leading-order term, this shows we need new physics below the scale $4\pi f_\theta$. In the presence of broken SUSY, we expect the UV divergence to be ameliorated, but this amounts to replacing the loop factors with $m_{\sigma}^2/(4\pi f_\theta)^2$ factors, and we conclude that $m_{\sigma} \lesssim 4\pi f_\theta$ is required for the loop expansion not to break down.

This conjecture is interesting as it suggests that periodic scalars never appear ``out of nowhere'' in a consistent theory; although a theory of a free compact boson is well-defined, in an interacting theory we always expect associated dynamics at or below the scale $4\pi f_\theta$. Unlike in Swampland bounds that crucially involve gravity, no factor of $M_{\rm Pl}$ appeared here. We expect that the radial mode mass bound applies to a much wider class of interacting theories, such as EFTs that approach (non-gravitational) conformal UV fixed points.

\bigskip
\noindent
{\em UV Cutoff Constraint for a Fundamental Axion}

If we find a quantum gravity theory with a {\em fundamental} axion of decay constant $f_\theta$, i.e.~one for which there is no point in field space at which the period of the axion shrinks to zero, we claim that the theory should have an ultraviolet cutoff
\be
\Lambda_{\rm UV} \lesssim \sqrt{f_\theta M_{\rm Pl}}.   \label{eq:axionUVcutoff}
\ee
This is a strong cutoff, in the sense that we expect local field theory to break down by this scale; we can't simply integrate in a finite number of weakly-coupled particles to fix it. In other words, this cutoff should be identified with the ``quantum gravity cutoff'' as discussed in e.g.~\cite{Heidenreich:2017sim}. We will motivate the particular power of $f_\theta$ appearing in the bound shortly.

\subsection{Towers of states and the ultraviolet cutoff}

For fundamental axions, as we have defined them, the Ooguri-Vafa Swampland Distance Conjecture tells us that a tower of fields becomes light as the axion period goes to zero. Suppose, for the sake of illustration, that this happens as follows. There is a tower of particles labeled by an integer $n$ such that 
\be
m_n \sim c n f_\theta^\gamma M_{\rm Pl}^{1-\gamma},
\ee
with $\gamma > 0$ some unknown order-one number. This has the necessary form for the Swampland conjecture: if we decouple gravity there is no constraint, whereas if we send $f_\theta \to 0$ the tower becomes massless.

Now we apply the species bound, which says that quantum gravity has a cutoff $\Lambda_{\rm UV} \lesssim M_{\rm Pl}/\sqrt{N}$ where $N$ is the number of weakly coupled modes below the scale $\Lambda_{\rm UV}$ \cite{ArkaniHamed:2005yv, Dvali:2007hz, Dvali:2007wp}. Counting particles in the tower, this is $N \sim \Lambda_{\rm UV}/(c f^\gamma M_{\rm Pl}^{1-\gamma})$. From this we can read off
\be
\Lambda_{\rm UV} \lesssim c^{1/3} M_{\rm Pl} \left(\frac{f_\theta}{M_{\rm Pl}}\right)^{\gamma/3},
\ee
which is just \eqref{eq:generalalpha} with $\alpha = \gamma/3$. This illustrates how the tower of states that the Ooguri-Vafa Swampland Distance Conjecture tells us must exist can be correlated with an ultraviolet cutoff at which local quantum field theory breaks down. This is the strongest statement that we can make without invoking a conjecture more precise than the Ooguri-Vafa conjecture. Note that the tower of states becoming light might have a more complicated spectrum; there could even be extended objects like strings that become light (and we will shortly argue that there are). In that case the counting of modes in the species bound becomes more subtle \cite{Dvali:2009ks, Dvali:2010vm, Heidenreich:2017sim}, though we expect the qualitative conclusion to still hold. In any case, we will now give a related, but sharper, argument based on the Weak Gravity Conjecture, which selects \eqref{eq:axionUVcutoff} as the precise form of our bound.

\subsection{An ultraviolet cutoff from the dual $B$-field WGC}
\label{subsec:Bfield}

A massless, compact axion field $\theta$ is Hodge dual to a massless 2-form gauge field $B = \frac{1}{2} B_{\mu \nu} \dif x^\mu \wedge \dif x^\nu$. Specifically, if $\theta$ has the periodicity $\theta \cong \theta + 2\pi$ and the Lagrangian\footnote{Throughout this paper I work in ($+$$-$$-$$-$) signature.}
\be
{\cal L}_\theta = \frac{1}{2} f_\theta^2 \partial_\mu \theta \partial^\mu \theta,
\ee
then the field strength $H = \dif B$ of the dual 2-form is given by $\frac{1}{f} H = f_\theta \star\!\dif\theta$ where the coupling constant $f$ of the $B$-field is $f = 2\pi f_\theta$:
\be
{\cal L}_B = \frac{1}{2 (3!) f^2} H_{\mu \nu \lambda} H^{\mu \nu \lambda}.
\ee
The factor of $2\pi$ relating $f$ and $f_\theta$ is the usual factor appearing in Dirac quantization, viewing $\theta$ as a zero-form gauge field with coupling $1/f_\theta$. The normalization of the field is chosen so that $B$ couples to unit-charge strings with worldsheet $\Sigma$ through the action $S = \int_\Sigma B$.

The WGC applied to the $B$-field implies the existence of low tension strings charged under $B$\cite{Arkanihamed:2006dz},
\be
T \lesssim f M_{\rm Pl}.      \label{eq:tensionbound}
\ee
This is a somewhat degenerate case of the WGC, as sufficiently high-tension strings coupled to gravity in $3+1$ dimensions have a deficit angle that destroys the space, so versions of the WGC based on the decay of extremal charged black objects do not directly apply. Nonetheless, there is evidence that the WGC continues to hold for 2-form gauge fields. Indeed, there is an argument based on black hole evaporation that applies precisely to this case. One can consider black holes which have arbitrarily large values of the axionic charge $b = \int_\Sigma B$, where $\Sigma$ is a 2-sphere homotopic to the black hole horizon \cite{Bowick:1988xh}. In the absence of charged strings, black hole evaporation maintains the value of $b$, leading to a remnant problem. Avoiding this pathology has been argued to imply the existence of strings charged under the $B$-field and obeying the bound \eqref{eq:tensionbound}~\cite{Hebecker:2017uix}.

We expect (and WGC arguments similarly imply) that the continuous shift symmetry of $\theta$ is always broken by instanton effects, which in nonsupersymmetric theories will generically give a mass to $\theta$. From the $B$-field point of view, this means that the charged strings are confined. Nonetheless, given that the instanton effects are generally exponentially suppressed, we expect that this does not modify the basic argument implying the existence of low-tension strings.

At this point it is important to distinguish between the cases of fundamental axions and ordinary Nambu-Goldstone bosons. We expect that \eqref{eq:tensionbound} applies in both cases. Theories of Nambu-Goldstone bosons generally admit semiclassical string solutions, and in a theory with a generic potential for the symmetry-breaking field, they will most likely have tension of order $f^2$. In the core of such a semiclassical string, the symmetry-breaking VEV goes to zero. Because the string can be entirely understood in terms of effective field theory, its existence implies no particular consequences for the ultraviolet behavior of the theory. The case of a fundamental axion is very different. In this case, there is no symmetry-restoring point at finite distance in field space. This means that the core of the string is singular. This is characteristic of strings that are fundamental objects (fundamental strings or $D$-strings, for instance, but objects associated with quantum gravity in any case). In this case, we interpret the energy scale $\sqrt{T}$ as an ultraviolet cutoff on local quantum field theory. Hence, for a {\em fundamental} axion, unlike a generic Nambu-Goldstone boson, we have a constraint
\be
\Lambda_{\rm UV} \lesssim \sqrt{f M_{\rm Pl}}.       \label{eq:BfieldUV}
\ee
Again, this argument was prefigured by~\cite{Hebecker:2017uix}. Below we will extrapolate the validity of this inequality to the case where a fundamental axion is eaten to generate a St\"uckelberg photon mass, providing the central claim of this paper.

\begin{figure}[tp]
\centering
\includegraphics [width = 0.7\textwidth]{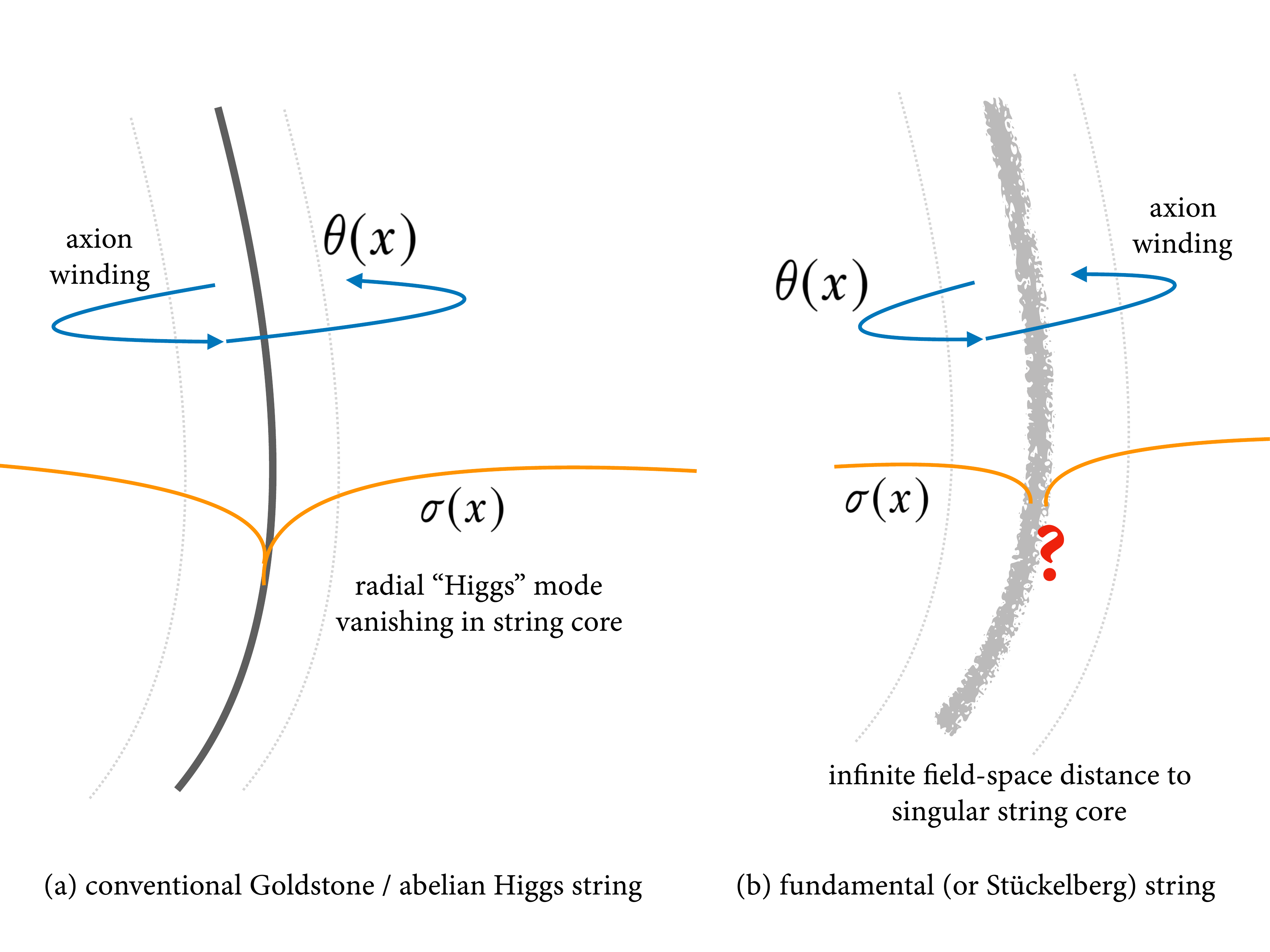}
\caption{
Illustrating the distinction between field-theoretic axion (or abelian Higgs model) strings and fundamental (or St\"uckelberg) strings. The string at long distances is very similar in the two cases. The difference is that a fundamental string---including those appeaing in the St\"uckelberg theory---has a singular core, as the would-be symmetry-restoring point lies an infinite distance away in field space. In the latter case, but not the former, we interpret the square root of the string tension as an upper bound on the energy at which local effective field theory is valid. 
}
\label{fig:strings}
\end{figure}

We give some explicit examples as further evidence for the validity of this bound in Appendix~\ref{app:examples}.

\section{St\"uckelberg masses in quantum gravity}
\label{sec:stuckelberg}

\subsection{Conjectures about St\"uckelberg masses}

By gauging the shift symmetry of compact boson $\theta$, we obtain the analogues for massive gauge fields of the conjectures of the previous sections. Specifically, we conjecture:

\bigskip
\noindent
{\em Compactness of the Gauge Groups}

A St\"uckelberg mass in quantum gravity arises from a U(1) gauge boson eating a compact scalar field. We will offer some comments below on whether this assumption can be relaxed, but it is borne out by many examples in string theory.

\bigskip
\noindent
{\em Radial (Higgs) Mode Mass Bound}

A massive gauge field should be accompanied by a radial ``Higgs'' mode $h(x)$ with an $h A_\mu A^\mu$ coupling. Just as in the radial mass bound for axions, we expect
\be
m_h \lesssim 4\pi f_\theta = 4\pi m_A/e.
\ee
We expect this to hold in both Higgs theories (where $m_A \to 0$ somewhere in field space) and in St\"uckelberg theories.

\bigskip
\noindent
{\em UV Cutoff of the St\"uckelberg Theory}

If we find a quantum gravity theory with a St\"uckelberg gauge boson with gauge coupling $e$ and mass $m_A = e f_\theta$, we claim that the theory should have an ultraviolet cutoff on the validity of local quantum field theory
\be
\Lambda_{\rm UV} \lesssim \min(e^{1/3} M_{\rm Pl}, \sqrt{f_\theta M_{\rm Pl}}) = \min(e^{1/3} M_{\rm Pl}, \sqrt{m_A M_{\rm Pl}/e}).
\ee
The $\sqrt{f_\theta M_{\rm Pl}}$ bound follows from viewing the gauge field as eating a fundamental axion and applying the UV cutoff conjecture of the previous section. The $e^{1/3} M_{\rm Pl}$ bound is the Sublattice or Tower WGC bound from loops of particles in the tower \cite{Heidenreich:2016aqi,Heidenreich:2017sim}.

\subsection{The BF-theory formulation and WGC argument}

Starting from the Proca Lagrangian for a massive gauge field, we can put the theory in St\"uckelberg form by treating $A_\mu$ as a gauge field and introducing a scalar field $\theta$ which shifts under a gauge transformation. A further reformulation is arrived at by dualizing the scalar $\theta$ to a 2-form gauge field $B_{\mu \nu}$, as discussed in \S\ref{subsec:Bfield}. We can thus write the theory of a photon mass arising from a $B \wedge F$ term, together with possible currents charged under both $A$ and $B$:
\be
{\cal L} = -\frac{1}{4 e^2} F_{\mu \nu}F^{\mu \nu} + \frac{1}{12 f^2} H_{\mu \nu \rho}H^{\mu \nu \rho} + \frac{n}{8\pi} \epsilon^{\mu\nu\rho\sigma}F_{\mu \nu} B_{\rho \sigma} + A_\mu J^\mu + B_{\mu \nu} \Sigma^{\mu \nu} + {\cal L}_{\rm inv}(F, H).  \label{eq:lagBF}
\ee
With these conventions, the photon mass is
\be
m_\gamma =\frac{n}{2\pi} e f.    \label{eq:photonmass}
\ee
If we assume that $A$ and $B$ each have a compact gauge group U(1), then the coefficient $n$ is quantized in $\mathbb{Z}$, and we have normalized the fields so that the conserved electromagnetic current $J^\mu$ and string current $\Sigma^{\mu \nu}$ have integrally quantized charges. The term ${\cal L}_{\rm inv}(F, H)$ denotes additional interaction terms in the Lagrangian built out of the gauge invariant field strengths, such as $(F_{\mu \nu}F^{\mu \nu})^2$.

We can now apply the Weak Gravity Conjecture to the BF-theory formulation of the St\"uckelberg mass. From \eqref{eq:photonmass} we see that to send $m_\gamma \to 0$ we must either send $e$ (the coupling of the gauge field $A$) or $f$ (the coupling of the gauge field $B$) to zero. In either limit, we run afoul of the WGC. One might wonder whether we should apply the WGC to massive gauge fields in the first place. In fact, this argument was already mentioned in \cite{Kaloper:2016fbr}, though they stopped short of drawing any conclusion due to such concerns. However, we already know that the WGC applies to massive gauge fields with Chern-Simons type masses in AdS$_3$ \cite{Montero:2016tif}, where it can be directly proven with arguments in the boundary CFT. Furthermore, there is some evidence that the WGC extends to arguments about Chern-Simons terms in more general contexts \cite{Montero:2017yja}. Finally, if the WGC can be proved in terms of black hole physics, consideration of black holes of radius much smaller than the scale $m_\gamma^{-1}$ will likely provide motivation for extending the argument to massive gauge fields. For all of these reasons, we believe our conjecture to be a modest extension of existing WGC conjectures (though it is fair to say that it does not logically follow from them).

In parallel to the case of Nambu-Goldstone bosons discussed in \S\ref{subsec:Bfield}, in the Higgs case we still expect to find strings of low tension in the small-$f$ limit, but they will be semiclassical strings of the familiar ANO type \cite{Abrikosov:1956sx, Nielsen:1973cs}, with the Higgs field going to zero in the string core. Hence their existence does not imply new UV physics. We only infer the upper bound $\Lambda_{\rm UV} \lesssim \sqrt{f_\theta M_{\rm Pl}}$ in the St\"uckelberg case, not the Higgs case.

\subsection{Giving up the compactness assumption}

\subsubsection{Black hole evaporation and noncompact massive gauge fields}

Suppose that we wanted to abandon the assumption of compact gauge fields. There are general black hole arguments against {\em massless} non-compact gauge fields \cite{Banks:2010zn}. So the limit where we send $m_\gamma \to 0$ should be sick. How close can we take it to zero before running into a problem? One estimate is to use the discharge rate of a black hole in a theory of a massive photon \cite{Mirbabayi:2013sva},
\be
\tau_{\rm dis} \sim \frac{1}{m_\gamma^2 R_{\rm BH}},
\ee
which applies when $m_\gamma \ll R_{\rm BH}^{-1}$. If this time is very long, then we can run the usual arguments that black holes charged under non-compact gauge symmetries produce problematic amounts of entropy, because the evaporation process is nearly insensitive to the photon mass. Assuming we don't venture too close to extremality, the lifetime of a black hole from Hawking evaporation is parametrically
\be
\tau_{\rm BH} \sim M_{\rm Pl}^2 R_{\rm BH}^3.
\ee
Then the condition $\tau_{\rm dis} \gg \tau_{\rm BH}$, which allows us to run the usual arguments without the black hole having time to notice that the gauge symmetry is not exact, is
\be
m_\gamma \ll \frac{1}{M_{\rm Pl} R_{\rm BH}^2}.
\ee
Only by violating this condition can we allow for noncompactness. In principle we should be able to talk about black holes with size $R_{\rm BH} \gtrsim \Lambda_{\rm UV}^{-1}$. Thus, even if we allow ourselves to talk about non-compact gauge fields in quantum gravity, we still reach the conclusion that we should require $\Lambda_{\rm UV} \lesssim \sqrt{m_\gamma M_{\rm Pl}}$. At small $e$, this is even a stronger bound than the one that we have conjectured assuming compact gauge fields!

\subsubsection{BF theory and charge quantization}

Another argument arises from considering the construction of Wilson loop or surface operators in the $B \wedge F$ theory with Lagrangian \eqref{eq:lagBF}. Given this Lagrangian, the equation of motion obtained by varying with respect to $A_\alpha$ is:
\be
\frac{1}{e^2} \partial_\mu F^{\mu \alpha} - \frac{n}{4\pi} \epsilon^{\alpha \mu \rho \sigma} \partial_\mu B_{\rho \sigma} + J^\alpha - 2 \partial_\mu \frac{\delta {\cal L}_{\rm inv}}{\delta F_{\mu \alpha}}  = 0.
\ee
From this we see that if there are no charged objects to supply a current $J^\alpha$, the dual field strength for the 1-form gauge field $\star H$ is a total derivative. Similarly, the equation of motion obtained by varying with respect to $B_{\alpha \beta}$ implies that without a string current $\Sigma^{\alpha \beta}$, $\star F$ is a total derivative.

This suggests the need for both particles charged under $A$ and strings charged under $B$. The reason is that if the field strength is a total derivative, we can construct well-defined 't Hooft loop (or surface) operators for {\em any} charge, not just integer charge.\footnote{Again, I thank Thomas Dumitrescu for a suggestion along these lines; any flaws in the reasoning as presented here are my own.} Having a continuum of well-defined operators in a theory of quantum gravity is problematic. For instance, Harlow has argued that factorizability demands that all Wilson line operators can be broken on charged objects obeying a WGC-like bound \cite{Harlow:2015lma}. If Wilson loops exist for charges that are not quantized, then no finite number of charged particles can allow for this. In short, strings charged under the $B$-field must exist in order to avoid the existence of a {\em continuum} of particles magnetically charged under the $A$-field, and vice versa. This suggests that a consistent theory, as in familiar examples, will in fact have quantized charges for both the $A$-field and the $B$-field, despite the fact that there is no massless gauge field in the spectrum.

\subsection{Evidence for the conjecture}

Classic examples of St\"uckelberg vector bosons in string theory have mass near the string scale and obey our conjecture. It has long been appreciated that St\"uckelberg masses can lie below the string scale due to small string coupling or large volume \cite{Antoniadis:2000ena, Kiritsis:2002aj, Antoniadis:2002cs}. Studies have been carried out of such light St\"uckelberg dark photons in string compactifications, due to their phenomenological interest \cite{Goodsell:2009xc, Cicoli:2011yh}. In examples, small dark photon masses are correlated with low string scales; for instance, \cite{Goodsell:2009xc} finds a class of dark photons with mass
\be
m_\gamma^2 \sim \frac{1}{g_s} \frac{M_{\rm string}^4}{M_{\rm Pl}^2}.
\ee
Local field theory breaks down at $M_{\rm string} \sim g_s^{1/4} \sqrt{m_\gamma M_{\rm Pl}} \lesssim \sqrt{f M_{\rm Pl}}$, in accordance with our expectations. In general, all of these stringy examples can be explicitly written in $B \wedge F$ form in terms of compact gauge fields, and contain objects of quantized charges. 

One point that could deserve further attention is how our bounds generalize to theories with a large number of St\"uckelberg gauge fields. It may be that our conjectures are reflected in the size of individual elements of a mass matrix, but that an eigenvalue can be anomalously light, allowing for a moderate evasion of our expectations. Such a possibility has been discussed in the context of the St\"uckelberg portal for massive gauge fields mediating interactions between the Standard Model and hidden sectors \cite{Feng:2014eja, Feng:2014cla}. 

\section{Phenomenological implications}
\label{sec:pheno}

Now that I have explained the conjectures and some reasons for believing them to be true, let us explore the phenomenological implications. We have seen that taking either $e$ or $f$ to be small yields an ultraviolet cutoff from WGC-like arguments. We can compromise by taking {\em both} $e$ and $f$ to be small. By taking $e^{1/3} M_{\rm Pl} \sim (m_\gamma M_{\rm Pl}/e)^{1/2}$, we find the weakest bound on the UV cutoff for a given St\"uckelberg mass:
\be
\Lambda_{\rm UV} \lesssim m_\gamma^{1/5} M_{\rm Pl}^{4/5}, \quad {\rm when} \quad e \sim \left(\frac{m_\gamma}{M_{\rm Pl}}\right)^{3/5}.     \label{eq:weakestbound}
\ee
As illustrated in Fig.~\ref{fig:UVcutoff}, this bound is often far below the Planck scale, and possibly below other scales of interest such as the Hubble scale during inflation.

\subsection{The Standard Model photon mass}

If our conjectures are true, the Standard Model photon mass must be exactly zero. The electromagnetic gauge coupling $e = \sqrt{4\pi \alpha} \approx 0.3$, so given the kinematic bound from FRBs $m_\gamma \lesssim 10^{-14}~{\rm eV}$, the St\"uckelberg UV cutoff bound implies that
\be
\Lambda_{\rm UV} \lesssim \sqrt{m_\gamma M_{\rm Pl}} \lesssim 5~{\rm MeV}, 
\ee 
which is impossible since we have tested local quantum field theory at higher energies. Furthermore, we know that there is no charged Higgs boson interacting with the photon with a mass below a TeV, so it is impossible for the Standard Model photon to get a mass from the Higgs mechanism. Indeed, even if the UV cutoff conjecture for St\"uckelberg masses is wrong, if the radial mode conjecture is correct, then the absence of any observed Higgs-like boson again rules out a photon mass.

Since some of my readers are phenomenological model-builders (as am I), they may share the impulse to be perverse and try to find a scenario consistent both with these conjectures and with observational data. Let me supply a loophole and the first steps toward such perversity. Suppose that we have misidentified the quantization of U(1) charge in the Standard Model, so that the $e$ that appears in all of our bounds (call it $e_0$) is not the measured $e$, but rather
\be
e_{\rm measured} = N e_0, \quad {\rm where} \quad N \gg 1.
\ee
In other words, perhaps the electron charge is not $-1$ (or, taking the down quark to be our unit of charge, $3$), but rather a trillion or some other huge integer. From \eqref{eq:weakestbound} we see that there is plenty of room for this to be potentially consistent: the largest UV cutoff allowed for $m_\gamma \sim 10^{-14}~{\rm eV}$ is at $10^{10}~{\rm GeV}$, well above the energy scales where we have experimental tests of local QFT. Not only that, but we could consider masses arising from the {\em Higgs} mechanism rather than the St\"uckelberg mechanism, so that the only UV cutoff is at $e^{1/3} M_{\rm Pl}$. In either case a light radial mode (an ``electromagnetic Higgs boson'') exists that couples to the photon, but if its charge is sufficiently small compared to the electron charge, we may not have observed it yet.

The electromagnetic Higgs boson would appear as a millicharged particle, and so is subject to many experimental constraints (see~\cite{Davidson:2000hf,Vogel:2013raa,Chang:2018rso} and references therein). However, these could be satisfied. For instance, we could have $f \sim {\rm eV}$ and $e_0 \sim 10^{-14}$ to achieve a photon mass of around $10^{-14}~{\rm eV}$. The electromagnetic Higgs boson would be at or below $10~{\rm eV}$, depending on its quartic coupling $\lambda$. This is roughly consistent with millicharged particle bounds requiring a charge $\epsilon \lesssim 10^{-14}$ for particles with mass below the keV scale.

This theory is perverse for two reasons. The first is the very large integer we have put in to account for the ratio of the electron charge to the electromagnetic Higgs boson charge. It seems to be difficult to find light particles with very large charges in string theory \cite{Ibanez:2017vfl} and there are some reasons to suspect that any attempt to do so will come at the cost of lowering the UV cutoff of the theory \cite{Heidenreich:2015wga}. Nonetheless, to the best of my knowledge no sharp argument exists ruling this out, and explaining the size of this integer could be a focus of model-building efforts. The second problem is of course that theories with a Higgs, or even theories of the radial mode that we expect to appear in the St\"uckelberg case, have a hierarchy problem. We could assume the Higgs has a tiny quartic coupling $\lambda_h \sim e_0^2$, but this will not be the only source of corrections. For instance, if the electromagnetic Higgs boson is elementary, then we expect graviton loops to raise its mass; in the SUSY context we estimate $\delta m_h \sim m_{3/2} \Lambda/M_{\rm Pl}$, with $\Lambda$ a UV cutoff. Putting in $\Lambda \gtrsim \sqrt{F} \sim \sqrt{m_{3/2} M_{\rm Pl}}$ we conclude that naturalness requires $\sqrt{F} \lesssim 40~{\rm TeV}$ if $m_h \sim m_\gamma \sim 10^{-14}~{\rm eV}$, which is difficult to achieve in a concrete model. However, one could eschew SUSY and generate the electromagnetic Higgs mass from strong dynamics, using a hidden confining gauge group with a low confinement scale. Since the photon couples so weakly to the hidden sector, this could be compatible with current constraints.

In short, the conjectures I have discussed strongly suggest that the Standard Model photon is exactly massless. Some work would be needed to close all the loopholes in the argument. Alternatively, building a UV-consistent theory that exploits the loopholes would be an interesting challenge that I encourage perverse readers to undertake. Similarly, experimentalists should be strongly encouraged to continue searching for signs of a nonzero photon mass (or of the less explored photon quartic coupling $\lambda_\gamma A_\mu^4$), which would either invalidate my conjectures or require the sort of unpleasant evasions I have just sketched.

\subsection{Dark photons}

``Dark photons,'' massive (but light) vector bosons, are a focus of intense experimental investigation. They may constitute dark matter themselves or mediate forces between dark matter particles. In general, a dark photon will kinetically mix with the Standard Model photon. The wide variety of strategies for searching for such particles has been reviewed in \cite{Essig:2013lka}. Dark photon masses near $10^{-20}~{\rm eV}$ could be interesting for the ``fuzzy dark matter'' scenario \cite{Hu:2000ke}, while dark photon masses at the $\mu{\rm eV}$ scale and above are interesting as they could constitute dark matter populated by inflationary fluctuations \cite{Graham:2015rva}. (For earlier discussions of dark photon dark matter and its cosmological abundance, see \cite{Nelson:2011sf, Arias:2012az}.) Inflationary fluctuations lead to the correct DM abundance provided that the dark photon has a St\"uckelberg mass
\begin{align}
m_\gamma &= 6~\mu {\rm eV}~\left(\frac{10^{14}~{\rm GeV}}{H_I}\right)^4, \label{eq:darkphotonrelic}
\end{align}
with $H_I$ the Hubble scale during inflation. The meaning of ``St\"uckelberg'' in this context is that the mass was turned on during inflation. This could happen in either the St\"uckelberg scenario as defined in this paper or in the case of a Higgs scenario where the Higgs boson is quite heavy; however, in either case, we expect that it will be difficult to decouple the radial mode so that the calculation is valid.
 
We can constrain the scenario assuming {\em either} the radial mode mass bound or the St\"uckelberg UV cutoff bound and using the requirement $\Lambda_{\rm UV} \gtrsim H_I$. Let us consider them one at a time. The assumption that the radial mode plays no role during inflation requires that $m_\sigma \gtrsim H_I$, which together with the radial mode mass bound leads to
\be
H_I \lesssim m_\sigma \lesssim 4\pi f.
\ee
This tells us that we are interested in a scenario where $e = m_\gamma / f \ll 1$. Then we apply the Tower or Sublattice WGC cutoff: 
\be
H_I \lesssim \Lambda_{\rm UV} \lesssim e^{1/3} M_{\rm Pl} \lesssim \left(\frac{4\pi m_\gamma}{H_I}\right)^{1/3} M_{\rm Pl}.
\ee
Comparing this to \eqref{eq:darkphotonrelic}, we see that obtaining the correct relic abundance from inflationary fluctuations requires 
\be
m_\gamma \gtrsim 60~{\rm eV}.
\ee
Hence, a substantial part of the parameter space for light dark photon dark matter from inflationary fluctuations, assuming a decoupled radial mode, is incompatible with the radial mode mass bound.

Next, we can ignore the radial mode mass bound but consider the St\"uckelberg UV cutoff bound. In this case we choose the coupling $e$ to attain the weakest bound \eqref{eq:weakestbound}. We have
\be
H_I \lesssim \Lambda_{\rm UV} \lesssim m_\gamma^{1/5} M_{\rm Pl}^{4/5},
\ee
which together with \eqref{eq:darkphotonrelic} implies that the correct relic abundance can only be attained when 
\be
m_\gamma \gtrsim 0.3~{\rm eV}.
\ee
Again, a large part of the parameter space is excluded by the bound.

These results show the potential power of Swampland considerations for constraining theories of physics beyond the Standard Model. Of course, they do not entirely exclude dark photon dark matter---only potential scenarios for obtaining the correct relic abundance. Other mechanisms of populating dark photon dark matter are possible \cite{Agrawal:2018vin}, though they too may be subject to interesting Swampland constraints.

\section{Further comments and clarifications}
\label{sec:comments}

Before concluding, I offer a few further remarks, mostly in response to questions and comments I have received since the first preprint version of this paper appeared.

First, there are questions about how to identify which massive spin-1 particles the conjectures apply to. If we take a theory containing a massless abelian gauge field and compactify on a circle, we obtain an infinite tower of massive spin-1 KK modes. The point at which they are massless is infinitely far away in field space. Does the conjecture apply? As another example in a similar spirit: suppose that we consider a $\rho$ meson of a confining gauge theory. If there is a limit in which the confinement scale goes to zero, does the conjecture apply to the $\rho$ meson? These examples are different from those I had in mind, in that they inherently involve a tower of particles of comparable mass which all become light in a uniform way (KK modes in the first case, hadrons in the second). By contrast, I have had in mind limits in which there is a single massive spin-1 field that can be parametrically separated from other modes. Because the cases of KK modes and $\rho$ mesons intrinsically involve towers of modes becoming light, I expect that some form of Swampland conjecture applies. But there is no reason to expect that a quantized coupling is identifiable or that a simple $B \wedge F$ formulation is useful in these contexts, so I do not believe my conjectures are directly applicable.

A related question involves mixing: there is no obstruction to mixing spin-1 fields obtaining mass from the Higgs mechanism and those obtaining St\"uckelberg masses. However, in this case I expect that there is still a locus in field space for which we obtain a massless mode, which will decompose into a subspace at finite distance and a subspace at infinite distance. So I see no fundamental difficulty in extending the arguments to the general context with mixing.

A final comment is on the nature of the argument about the gravitational cutoff $\Lambda_{\rm UV}$. The form of the WGC that I have invoked demands strings of bounded tension $T$, which must be fundamental in the sense that the EFT breaks down in the core of the string. I have interpreted the scale $\sqrt{T}$ as a fundamental UV cutoff, above which no local effective quantum field theory involving a finite number of modes is valid. This interpretation may be too hasty. One could object that if the strings are sufficiently weakly coupled, their existence may not actually affect generic particle physics processes. Perhaps there are MeV-tension strings in our universe which are fundamental (in the sense that EFT breaks down in the string core), but we simply don't interact with them with sufficient strength to have noticed. To this I have two tentative counterarguments, neither quite as clean as I would like. The first is that these strings have a spectrum of excited states, and that counting the states in this tower implies a fundamental cutoff near the string scale. It is tempting to say that the states will have a Hagedorn spectrum so that the species bound leads one to expect that the fundamental cutoff $\Lambda_{\rm UV}$ can be at most a factor of order $\log(M_{\rm Pl}/\sqrt{T})$ above the string tension scale $\sqrt{T}$. However, once the states appearing in the species bound become extended objects, it is not entirely obvious how to count them; see \cite{Dvali:2009ks, Dvali:2010vm} for an argument that the number of species in 10d string theory is $1/g_s^2$, which would lead to the identification of $\Lambda_{\rm UV}$ with $M_{\rm string}$. The second counterargument is that the application of the Swampland Distance Conjecture independently suggested a quantum gravity cutoff of the form $f_\theta^\alpha M_{\rm Pl}^{1-\alpha}$; however, this is quantitatively less powerful since it doesn't determine the constant $\alpha$. This leaves room for the bound to be somewhat weaker, if a theory can be exhibited in which the existence of strings that are fundamental (in the sense that I have used the term here, more generally than just the F-strings of string theory) does not imply a fundamental cutoff on the validity of EFT near the scale of the string tension. Any such example would be fascinating to study further.

\section{Conclusions}
\label{sec:conclusions}

I have argued that theories of quantum gravity, unlike effective field theory, impose strong constraints on the nature of spin-1 massive bosons. While small masses from the Higgs mechanism are possible, small masses of St\"uckelberg type imply a low ultraviolet cutoff. Like most Swampland arguments, these claims rest on conjectures. However, they are tightly linked to the most well-supported of the previously-studied Swampland conjectures, the Swampland Distance Conjecture and the Weak Gravity Conjecture, and are supported by similar evidence from concrete examples.

One route toward future progress could come from proofs in conformal field theory of statements that are dual to Swampland conjectures in AdS. For example, the conjecture about the mass of the radial mode implies the existence of a scalar single-trace operator that has a three-point function with two currents. We are interested in currents that are not conserved (dual to massive vector bosons), with dimension $1 \ll \Delta \ll \Delta_{\rm gap}$ where $\Delta_{\rm gap}$ is the CFT analogue of the UV cutoff scale. One might hope to prove such statements directly in CFTs.

There is a very active experimental program aimed at searching for dark photons. The arguments in this paper give further motivation to such work. The discovery of an ultralight spin-1 boson would require us to understand the nature of its mass, and could either support or refute conjectures about quantum gravity. 

The arguments in this paper cast some doubt on the wisdom of too strict an adherence to the notion of ``technical naturalness.'' Tiny St\"uckelberg masses can be added to an effective field theory at no cost, and are radiatively stable. However, in quantum gravity they could be much more dangerous, implying a breakdown of local QFT at low energies. The same may be more generally true of tiny parameters in effective field theories that are not explained in terms of any underlying dynamical mechanism. The fuzzier notion of a natural theory as one in which all the parameters are explained in terms of order-one numbers and dynamics may, ultimately, be more robust than apparently sharper questions about radiative stability.

\section*{Acknowledgments}

I thank Thomas Dumitrescu for collaboration at an early stage of this work, Prateek Agrawal and Arvind Rajaraman for useful discussions, JiJi Fan for feedback on the draft, and Elias Kiritsis for informing me about the link between 4d St\"uckelberg masses and 6d anomalies in string theory. I also thank Brando Bellazzini, Ben Heidenreich, Elias Kiritsis, and Tom Rudelius for useful feedback on the first preprint version. I thank the students of Harvard's Physics 253a in Fall 2017 for helping me to appreciate how annoying it is that quantum field theory doesn't give us a compelling reason to believe the photon is exactly massless. My work is supported in part by the DOE Grant DE-SC0013607 and the NASA ATP Grant NNX16AI12G.

\appendix

\section{Is dynamical symmetry breaking a special case of Higgsing?}
\label{app:DSB}

We have defined the crucial distinction between the St\"uckelberg mechanism and Higgsing to be whether the zero-mass point in field space lies at finite or infinite distance. In the introduction, I claimed that dynamical symmetry breaking, for instance by QCD-like strong dynamics, should be thought of as Higgsing. Is this strictly correct? Strong dynamics generates low masses by dimensional transmutation, so that our confinement scale behaves as 
\be
\Lambda_{\rm conf} \approx M \E^{-\frac{8 \pi^2}{c g^2(M)}},
\ee
where $c$ is some constant depending on the number of colors and flavors of the theory and we have assumed a weak gauge coupling $g^2(M)$ at an energy $M \gg \Lambda_{\rm conf}$. We can naturally achieve an exponentially small scale of symmetry breaking using a mildly small value of $g^2(M)$. Nonetheless, $g^2$ is a gauge coupling and as such the $g \to 0$ limit in quantum gravity lies at infinite distance in field space (which we expect to diverge as $|\log(g)|$ \cite{Ooguri:2006in, Heidenreich:2018kpg, Grimm:2018ohb}). This infinite distance is, as always, associated with an ultraviolet cutoff. For example, in SU(2) gauge theory the Sublattice WGC implies that the quantum gravity cutoff energy is at or below $g^{1/2} M_{\rm Pl}$ \cite{Heidenreich:2017sim}.

This suggests that, as with the St\"uckelberg mechanism, obtaining a small vector boson mass from dynamical symmetry breaking will imply a cutoff below the Planck scale, and that the $m_A \sim e \Lambda_{\rm conf} \to 0$ limit will bring the cutoff arbitrarily low. However, the functional form of the cutoff is {\em extremely} different from what we have derived in the usual St\"uckelberg case. Taking the SU(2) case as an example, we have a quantum gravity scale
\be
\Lambda_{\rm UV} \lesssim g(M)^{1/2} M_{\rm Pl} \sim \left(\frac{8 \pi^2}{c \log(M/\Lambda_{\rm conf})}\right)^{1/4} M_{\rm Pl}.
\ee
Due to its extremely mild logarithmic dependence on $\Lambda_{\rm conf}$, this is not a useful bound (even if we were to take a confinement scale of order Hubble). Hence the identification of dynamical symmetry breaking with the Higgs case, rather than the St\"uckelberg case, is the correct one for all practical purposes even if it is strictly speaking incorrect. We could reformulate the distinction to say that our arguments against light St\"uckelberg fields apply to those theories in which the field-space distance to a point with a vector mass $m_A$ grows at least as fast as $|\log(m_A^2)|$; the strong dynamics case has a field space distance growing rather as $\log|\log(m_A^2)|$, implying much weaker constraints.

\section{Examples supporting the fundamental axion UV cutoff argument}
\label{app:examples}

In \S\ref{sec:axion} we have argued that theories of fundamental axions with a small decay constant $f$ have an associated low UV cutoff \eqref{eq:BfieldUV}. In this appendix we give some further evidence for this claim.

\subsection{Simple examples in string theory}

Let us review some explicit examples for concreteness. Suppose that we compactify 10d string theory on a manifold with volume $\cal V$ in units of the string length $\ell_s$. Then the 4d Planck scale is determined by (neglecting order-one factors) $M_{\rm Pl}^2 \sim \frac{1}{g_s^2 \ell_s^2} {\cal V}.$ The $B$-field in 10 dimensions has action $\sim \int \dif{^{10}x}\, \frac{1}{g_s^2 \ell_s^4} |\dif B|^2$. The prefactor in the 4d kinetic term of the $B$-field is then $\frac{1}{f^2} \sim \frac{{\cal V} \ell_s^2}{g_s^2}.$ At either small coupling $g_s$ or large volume $\cal V$, the 4d $B$-field is weakly coupled and its dual axion has a low decay constant $f \ll \ell_s^{-1}$. As expected from the Swampland Distance Conjecture, sending either $g_s \to 0$ or $\cal V \to \infty$ brings down a tower of fields: string excitations in the former case and Kaluza-Klein modes in the latter. In this example \eqref{eq:BfieldUV} translates to $\Lambda_{\rm UV} \lesssim \sqrt{f M_{\rm Pl}} \sim \ell_s^{-1}$,  precisely the string scale at which local quantum field theory breaks down. Next, consider a case where a $p$-form Ramond-Ramond field $C_p$ in 10 dimensions leads to a 2-form field $C$ in 4d by integration over a $p-2$ cycle of volume ${\cal V}_{p-2}$ in string units. In that case the 10d action is $\sim \int \dif{^{10}x}\, \frac{1}{\ell_s^{8-2p}} |\dif C_p|^2$ and the 10d charged objects are $(p-1)$-branes of tension $T_{p-1} \sim \frac{1}{g_s \ell_s^p}$. Reducing to 4d, we have $C \sim {\cal V}_{p-2} \ell_s^{p-2} C_p$ and so $1/f^2 \sim ({\cal V}/{\cal V}_{p-2}^2) \ell_s^2$. The tension of the charged strings arising from wrapped $(p-1)$-branes is $T \sim T_p {\cal V}_{p-2} \ell_s^{p-2} \sim \frac{1}{g_s} {\cal V}_{p-2} \ell_s^{-2}$. Because the strings originate in $D$-branes, their tension lies above the fundamental string scale. In this case, the WGC bound is again saturated: $T \sim f M_{\rm Pl}$, so if we identify $\Lambda_{\rm UV}$ with $\ell_s^{-1}$ we have $\Lambda_{\rm UV} \ll \sqrt{f M_{\rm Pl}}$ at small $g_s$ and large ${\cal V}_{p-2}$ where our calculations are under control. If ${\cal V}_{p-2} \ll g_s$, the strings arising from wrapped branes become even lighter than fundamental strings. If this conclusion is accurate (which is unclear, since our approximations could be altered by higher derivative terms in the small-volume limit), it seems that the appropriate conclusion is that we should then interpret {\em their} tension scale, rather than $\ell_s^{-1}$, as the cutoff on local field theory.

\subsection{Conformal collider bounds}

An additional set of examples may be found in conformal field theories dual to quantum gravity theories in AdS. A generic axion $a$ with period $2\pi f$ can have a nonminimal coupling to gravity of the form
\be
\frac{n}{32\pi^2 f} \int d^4 x\, a W \widetilde{W},
\ee
with $W$ the Weyl tensor, for integer $n$. There is a bound in the CFT context specifically in theories with a large $N$ expansion (i.e.~a clear notion of single-trace versus multi-trace operators) and a relatively small number of single-trace operators below dimension $\Delta_{\rm gap}$, which suggests that when such a coupling arises in an AdS spacetime, we must have \cite{Afkhami-Jeddi:2018own}
\be
\frac{n}{f} \lesssim \frac{M_{\rm Pl}}{M_{\rm gap}^2}.
\ee
Here $M_{\rm gap} = \Delta_{\rm gap}/\ell_{\rm AdS}$ is the scale at which large numbers of single-trace operators proliferate, roughly dual to the string scale. This implies that
\be
M_{\rm gap} \lesssim \sqrt{f M_{\rm Pl}/n}.
\ee
This looks quite similar to our proposed bound \eqref{eq:axionUVcutoff}.

This is not a very strong argument; a given axion could simply have $n = 0$, and in any case the CFTs in which this bound can be proven have special properties that distinguish them from generic quantum gravity theories in AdS. (A general theory in AdS may have many couplings of different sizes, rather than a uniform notion of a $1/N$ expansion.) Nonetheless, I find this result intriguing as a hint of how improved understanding of conformal field theory could connect with the study of quantum gravity in general and the Swampland program in particular.

%{\small
\bibliography{ref}
\bibliographystyle{utphys}
%}

\end{document}